\documentclass[runningheads]{llncs}
\usepackage[T1]{fontenc}

\usepackage{amsmath,amssymb,amsfonts}

\usepackage{graphicx}
\usepackage{float}
\usepackage{subcaption}
\usepackage[justification=centering, skip=2pt]{caption}

\usepackage{tikz}
\newcommand*\circled[1]{\tikz[baseline=(char.base)]{
            \node[shape=circle,draw,inner sep=1pt] (char) {#1};}}

\usepackage{booktabs}
\usepackage{array}

\usepackage{enumitem}

\usepackage{xcolor}

\usepackage{hyperref}

\usepackage{cleveref}

\def\BibTeX{{\rm B\kern-.05em{\sc i\kern-.025em b}\kern-.08em
    T\kern-.1667em\lower.7ex\hbox{E}\kern-.125emX}}

\usepackage{color,soul}
\usepackage{hyperref}
\usepackage{subcaption}
\usepackage{graphicx}
\usepackage{orcidlink}

\begin{document}
\title{ANVIL: Analogies and Videos for Lecturers}
\titlerunning{ANVIL}

\author{
Yuri Noviello\inst{1}\orcidlink{0009-0006-5846-7756} \and
Anastasiia Birillo\inst{2}\orcidlink{0000-0003-2269-8211} \and
Gosia Migut\inst{1}\orcidlink{0000-0002-4120-5454}
}
\authorrunning{Y. Noviello et al.}
\institute{
    Delft University of Technology, Delft, The Netherlands\\
    \email{
        \{y.noviello,m.a.migut\}@tudelft.nl  
        }\\
\and
    JetBrains Research,
    Belgrade, Serbia\\
    \email{
    anastasia.birillo@jetbrains.com
    }
}

\maketitle
\begin{abstract}
We present \textit{ANVIL}, a multimodal generative system that automates the production of analogy-based instructional animations for computer science topics. Given a concept definition, \textit{ANVIL} generates a textual analogy, compiles it into a structured visual screenplay, and produces executable \texttt{manim} code to render an animation, with an automated repair mechanism to improve robustness.
Evaluating such systems at scale requires balancing pedagogical validity with scalability. We begin with a teacher evaluation to ground the quality assessment and use its findings to guide automated screening. For textual analogies, we introduce an LLM-based evaluator for scalable quality screening; for videos, where subjective judgments are difficult to automate, we instead assess fidelity to the intended screenplay using an automated proxy for auditing and error analysis.
We further conduct a user study with educators to examine adoption requirements and risks.
Our findings suggest that \textit{ANVIL} can produce materials that are frequently rated as adequate, and that educators respond positively to its perceived value and usability.
\end{abstract}
   

\setcounter{footnote}{0}
\section{Introduction}
\label{sec:introduction}

Traditional educational methods often struggle to engage learners and  convey abstract concepts effectively~\cite{freeman_active_2014}, particularly in Computer Science (CS) and Software Engineering (SE), where topics such as algorithms and data structures, require high-level abstract reasoning.
A well-established approach to addressing this challenge is the use of analogies to map abstract concepts onto familiar domains~\cite{gentner_reasoning_1997}, supporting comprehension, recall, and engagement~\cite{ortony_mechanisms_1989}. When combined with multimodal visual explanations, analogies can further enhance conceptual understanding and motivation~\cite{bouchey_multimodal_2021}.
However, creating high-quality multimodal analogy-based educational materials remains resource-intensive and requires substantial pedagogical and multimedia expertise~\cite{snelson_quest-based_2022}.
Recent advances in Artificial Intelligence (AI), particularly in Large Language Models (LLMs), offer new possibilities for automating scalable educational content generation~\cite{mittal_comprehensive_2024}, including the systematic creation of instructional analogies and visualizations for CS education.

In this paper, we present \textbf{ANVIL} (\textbf{AN}alogies and \textbf{VI}deos for \textbf{L}ecturers), a multimodal generative system that automates the end-to-end production of multimodal, analogy-based instructional content for CS concepts.
\textit{ANVIL} employs a staged pipeline that generates an analogy (\textit{Textual Layer}), compiles it into a structured visual screenplay (\textit{Screenplay Layer}), and generates executable Python \texttt{manim} code (\textit{Code Layer}).
To improve robustness, ANVIL integrates a self-repair loop that combines static analysis and iterative automatic repair to recover from code-generation failures without human intervention.

Such a system must be thoroughly evaluated before it can be deployed in teaching~\cite{unesco}.
While expert judgment remains essential for assessing the instructional adequacy of analogies and animations, it does not scale to large numbers of artifacts.
We therefore adopt an evaluation approach in which teacher assessment serves as an initial grounding step, establishing how generated analogies and animations are perceived and revealing where reliable automation is feasible.
These findings guide the design of automated evaluation methods with modality-specific roles: 
(i) for analogies we investigate scalable screening via an LLM-based evaluator validated against expert judgments;
(ii) for animations we evaluate whether rendered animations faithfully realize their intended intermediate representations through automated screenplay-to-video \textit{fidelity proxy}. 

This framing treats automated evaluation as a mechanism for pre-selection and auditing, rather than as a substitute for expert review.
In addition, we conducted two focus groups with educators to examine requirements and guardrails for classroom deployment. These discussions focused on expectations around instructor control, editability, and risk mitigation when AI-generated materials may introduce misconceptions.
This paper makes four contributions:
\begin{enumerate}
    \item \textbf{System:} \textit{ANVIL}, an open-source system that generates analogical animations via intermediate representations, including an agentic repair mechanism to increase robustness.
    \item \textbf{Human evaluation:} A rubric-based evaluation of both generated analogies and rendered animations with 11 CS/SE educators.
    \item \textbf{Automated Evaluation:} An LLM-based analogy evaluator \emph{validated} against expert judgments, and a screenplay-to-video \emph{fidelity proxy} used for screening and error analysis.
    \item \textbf{User study:} A qualitative user study with $9$ CS/SE educators, characterizing perceived usefulness, risks, and integration requirements for deploying analogy-to-video generation in teaching workflows.
\end{enumerate}

\section{Related Work}
\label{Context_Related}
Analogies are a well-established instructional strategy for supporting understanding of abstract concepts by mapping them to familiar domains, with strong foundations in cognitive and educational theory~\cite{pramling_learning_2015,gentner_reasoning_1997}. Prior work in CS education shows that analogy-based instruction can reduce cognitive load and improve conceptual understanding, particularly for novice learners~\cite{saxena_achieving_2023,harsley_incorporating_2016}. However, effective analogies are typically handcrafted by instructors, requiring substantial pedagogical and domain expertise, resulting in limiting scalability~\cite{snelson_quest-based_2022}.

Recent work shows LLMs can assist students in crafting analogies on demand~\cite{bhavya_analogy_2022,bernstein_like_2024-1,jiayang_storyanalogy_2023,hu_-context_2023}.
Bhavya et al.~\cite{bhavya_analogy_2022} released a benchmark of analogy prompts with the correlated human judgments, and Bernstein et al.~\cite{bernstein_like_2024-1} demonstrated that LLMs could produce engaging and memorable analogies for recursion, especially when prompted with personal context.
While these results are promising, the reliability of LLM-generated analogies for educational purposes remains a key challenge.
Shao et al.~\cite{shao_unlocking_2025} found that LLMs may produce flawed analogies, violating commonsense relations or introducing inconsistencies.
We address this risk by designing \textit{ANVIL} around a staged generation pipeline.
Educators can review and revise the analogy before rendering, avoiding that incorrect analogies propagate into instructional materials.

Visualizations have long been used to teach computing concepts. Tools like
Online Python Tutor~\cite{guo_online_2013-1} and Jeliot~\cite{moreno_visualizing_2004}, effectively animate program execution to aid student comprehension. However, these tools are designed to visualize the concrete execution of code, not the abstract models that analogies can provide.
The \texttt{manim}~\cite{the_manim_community_developers_manim_2025} Python library has been widely adopted to create explanatory animations for computing courses, but as noted by Marković and Kastelan~\cite{markovic_demonstrating_2024}, producing high-quality content requires considerable authoring effort and multimedia expertise. Furthermore, a systematic review by Sibia et al.~\cite{aurisano_exploring_2025} highlights that most visualization tools focus narrowly on specific content (e.g., particular data structures or algorithms), limiting their flexibility. 
\textit{ANVIL} contributes by automating the \texttt{manim} authoring process, making video animations accessible to educators without specialized multimedia expertise. Moreover, our approach is topic-agnostic, allowing visualizations for a wide range of concepts.

Advancements in LLMs have inspired research into automatically generating visual content from natural language.
The system developed by Sehgal et al.~\cite{sehgal_exploring_2024} combines LLM-generated analogies with static images produced by text-to-image models.
While they demonstrate the potential of fusing visual elements with analogical explanations for science education, their focus remains on generating static images.
For many CS topics, dynamic animations can be useful for explaining processes, sequences, and state changes~\cite{markovic_demonstrating_2024}.

Community tools have begun to explore LLM-to-\texttt{manim} generation (\textit{e.g.}, the \emph{generative-manim} repository~\cite{arias_marcelo-earthgenerative-manim_2025}), reflecting growing practitioner interest. 
Concurrently, Manimator~\cite{p_manimator_2025} and Code2Video~\cite{chen2025code2video} leverage LLMs to generate \texttt{manim} animations, automating the rendering of mathematical concepts.
Our work adapts this technology specifically for CS education, with a focus on generating animations of conceptual analogies rather than only mathematical or algorithmic representations.
By integrating both analogy generation and dynamic animation into a single workflow, \textit{ANVIL} provides a novel end-to-end solution for creating instructional materials.
\begin{figure}[ht]
    \centering
    \includegraphics[width=\textwidth]{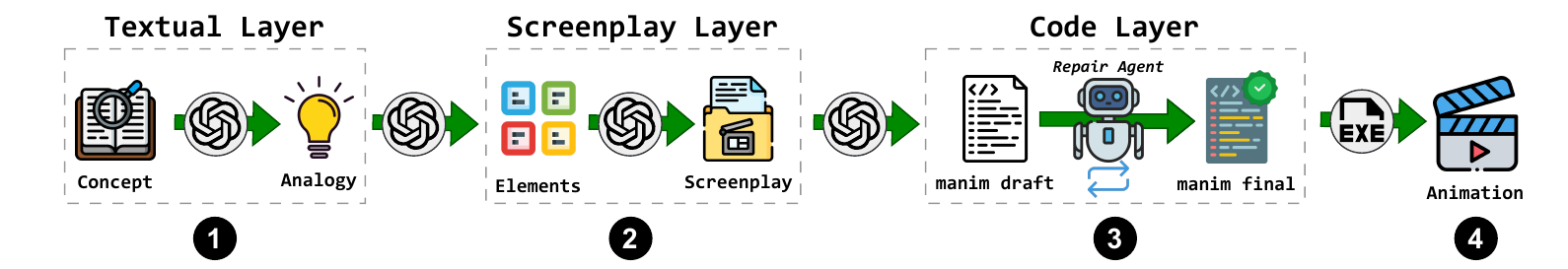}
    \caption{\small Overview of \textit{ANVIL}'s generation pipeline. Given a target CS concept, ANVIL (1) synthesizes an analogy, (2) compiles it into visual elements and a structured screenplay, (3) generates an executable \texttt{manim} program and applies a bounded agentic repair loop to fix compilation/runtime errors, and (4) renders the animation.}
    \label{fig:pipeline}
\end{figure}

\section{ANVIL Architecture}
\label{ANVIL_architecture}
\textit{ANVIL} integrates analogy generation and automatic creation of \texttt{manim}~\cite{the_manim_community_developers_manim_2025} animations using LLMs.
This section describes how the system performs these actions. \Cref{fig:pipeline} provides an overview of the system workflow.
The code and the prompts of the system are available in the paper's repository~\footnote{\url{https://github.com/yurinoviello/AI_Metaphors}}.

\subsection{Textual Layer}
\label{conceptual_framework}
The first stage converts a concept definition into an analogous explanation that can later be visualized.
We frame this action as a problem of \textit{mapping analogical domains}: a target domain $T$ (the CS concept) is explained through a source domain $S$ (a familiar scenario).
Given a concept and its definition, \textit{ANVIL} prompts an LLM to propose a candidate source domain $S$ and to construct an explicit mapping between the definition's properties and elements in the source scenario (\Cref{fig:pipeline}~\circled{1}).
To encourage pedagogically coherent analogies, we ground the prompt in Structure-Mapping Theory~\cite{mapping-theory} and introduce a coverage constraint in the generation instructions: for each property stated in the definition, the model must produce a corresponding property in the source domain.
The result of this layer is an analogy that explains the input concept.

\subsection{Screenplay Layer}
The goal of the Screenplay Layer is to translate the analogy into an executable visual plan.
In early experiments, the direct conversion from the analogy to the \texttt{manim} script proved unreliable, resulting in consistent issues like object overlap and inconsistent narrative structure.
To address this problem, we rely on two intermediate representations: (i) a set of \textit{visual elements} (objects/characters) expressed as \texttt{manim} class definitions, and (ii) a scene-level \emph{screenplay} that specifies object placements, state changes, and on-screen text over time~(\Cref{fig:pipeline}~\circled{2}).
~\\
\textbf{Elements Definition.} Given the analogy produced in the previous layer, \textit{ANVIL} decomposes the source scenario into a finite set of concrete visual elements and their expected behaviors. 
Each element is encoded as a structured record with a \texttt{name}, \texttt{role}, \texttt{actions}, and a \texttt{manim} class template. 
We prompt an LLM with instructions to:
(i) identify the key elements in the analogy, 
(ii) specify each element role,
(iii) enumerate the element's actions as animatable operations, and
(iv) synthesize a \texttt{manim} class definition for visualizing and manipulating the element.
To improve visual consistency across animations, \textit{ANVIL} provides the LLM with access to a curated catalog of vector graphics (\textit{SVG} assets). 
The prompt includes the list of available asset filenames, enabling the model to either (a) reuse an existing SVG (e.g., \texttt{doll.svg}, \texttt{tree.svg}) or (b) generate the element procedurally using \texttt{manim} primitives. 
The catalog is extensible: educators can add or remove SVG files to customize the visual assets.
~\\
\textbf{Screenplay Generation.} After elements have been defined, \textit{ANVIL} generates a natural-language \textit{screenplay} that specifies the animation as an ordered sequence of scenes, acting as an intermediate representation between the textual analogy and the final video animation.
Each scene describes (i) which elements are present, (ii) where they appear on the screen, (iii) what actions occur (including temporal order and state changes), and (iv) what explanatory text should be displayed.
The screenplay is generated using the elements generated previously, which constrains the set of allowed objects to prevent the hallucinations of undefined visual elements.

\subsection{Code Layer}
\label{sec:code_layer}

\textbf{Script Generation.}
The Code Layer combines the generated \emph{elements} and \emph{screenplay} into a self-contained \texttt{manim} program (\Cref{fig:pipeline}~\circled{3}).
We prompt an LLM with the full context (elements and screenplay) and a fixed template, including reusable utility functions to encourage consistent layout and transitions.
~\\
\textbf{Agentic Repair.}
LLM-generated \texttt{manim} code can fail due to API hallucinations, syntax errors, or inconsistent object references. 
To improve robustness, \textit{ANVIL} applies a bounded diagnose--repair--verify loop.
At each iteration, the system (i) runs static checks (via \texttt{pylint}~\cite{pylint_contributors_pylint_2025}) and, if needed, executes the script to find runtime errors, (ii) provides the resulting diagnostics together with the current code to a code-repair LLM, and (iii) re-runs the checks on the revised output.
We cap this loop at $n{=}3$ iterations. Repair is invoked only if errors are detected.
~\\
\textbf{Rendering.}
Finally, the repaired script is executed to render the analogical animation video (\Cref{fig:pipeline}~\circled{4}).
All intermediate artifacts (analogy, elements, screenplay, \texttt{manim} scripts, and the final video) are saved for inspection and reuse.

\section{Robustness Analysis}
\label{sec:repair_stats}

To quantify \textit{ANVIL}'s robustness in code generation, we measure how often the generations required automatic repair and how many iterations are needed before a runnable \texttt{manim} script is obtained.
Across $n{=}50$ end-to-end generation runs, we used \texttt{gpt-4o} for the Textual and Screenplay Layers and \texttt{claude-3.7-sonnet} for the Code Layer.
Overall, $38$ runs (76\%) succeeded without invoking repair.
The remaining $12$ runs required at least one repair iteration: $9$ runs (18\%) required a single iteration, $2$ runs (4\%) required two iterations, and $1$ run (2\%) required three iterations.
All runs completed within the preset repair-iteration limit, indicating that the agent reliably resolves common code-generation failures. Without the repair agent, 24\% of runs would have failed to render a video.

\section{Evaluation}
\label{Evaluation}
The \textbf{goal} of our evaluation is (1) to determine the quality of generated analogies and animations as judged by the expert teachers; and given the implications of human evaluation (2) to establish how the analogies and animations can be automatically screened for quality at scale without compromising pedagogical validity.
All evaluation materials are included in the replication package~\footnote{Replication package: \url{https://osf.io/h7gy6}}.

\begin{figure}[t!]
    \centering
    \includegraphics[width=0.49\linewidth]{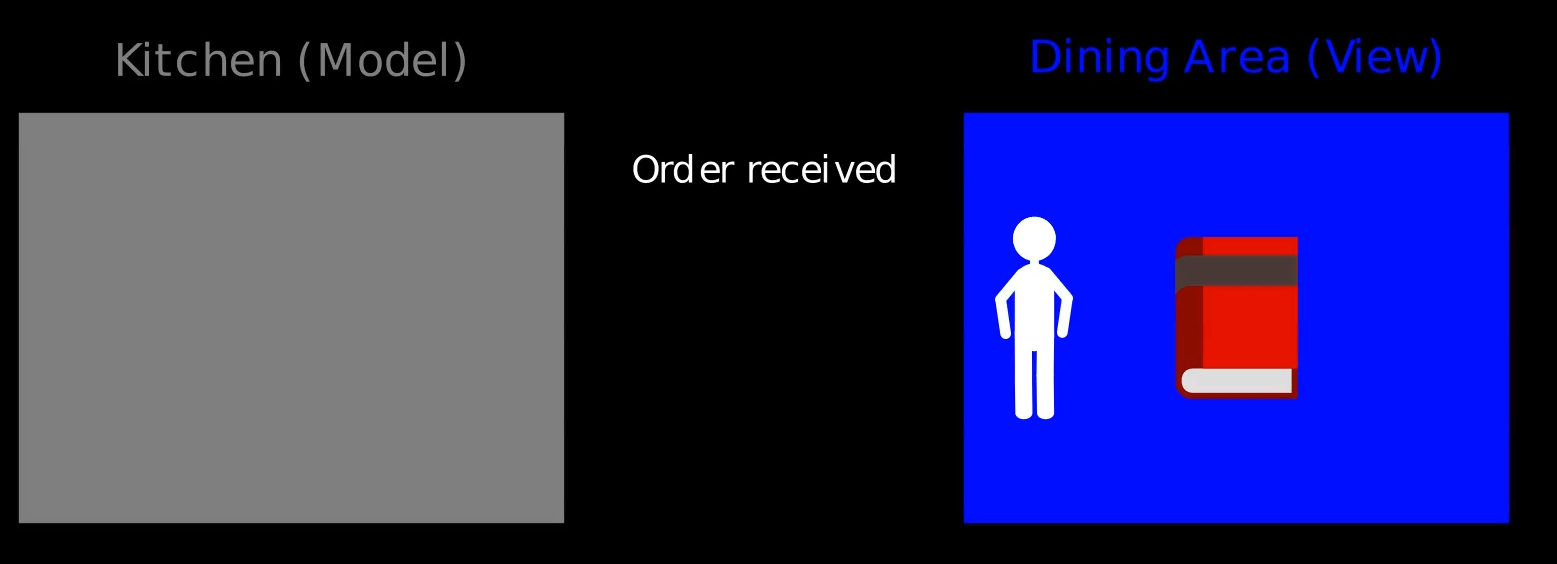}
    \includegraphics[width=0.49\linewidth]{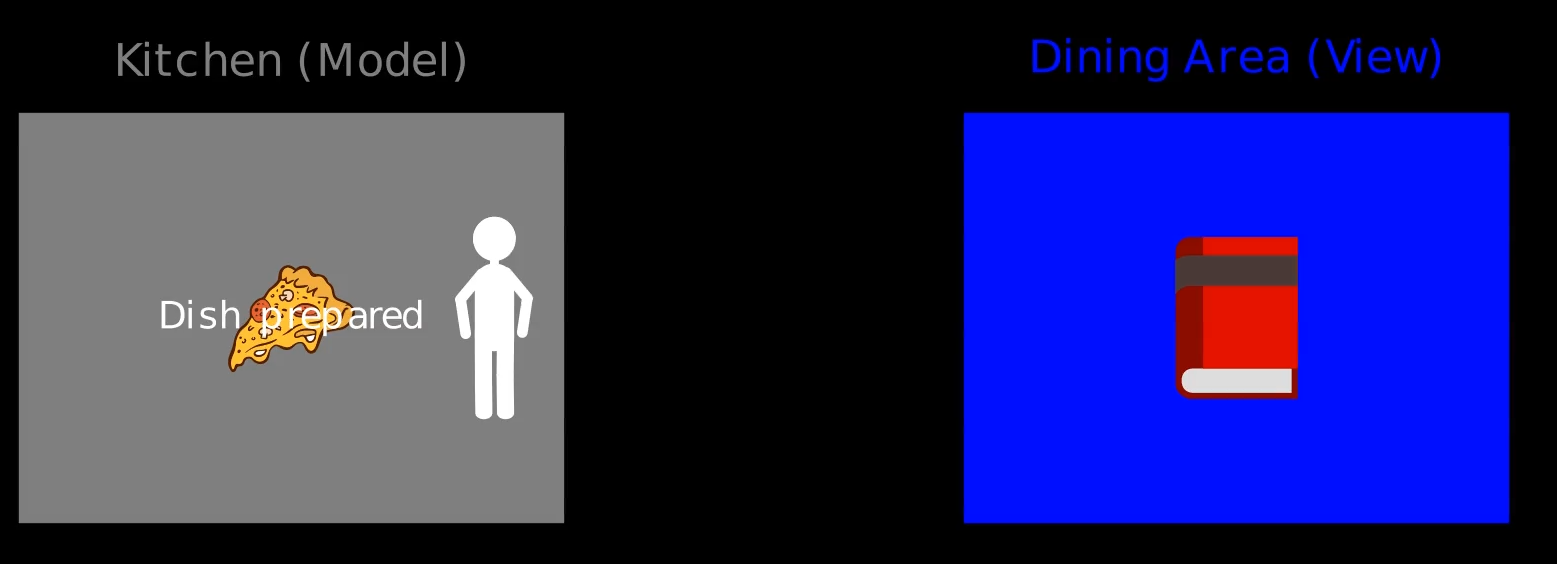}
\caption{\small Model--View--Controller Pattern as a Restaurant.
The View captures the \textit{order}. The Controller (\textit{waiter}) forwards it to the Model (\textit{kitchen}). The Model processes the order (\textit{dish prepared}).}
\label{fig:MVC}
\end{figure}

\subsection{Human Evaluation}
\label{sec:expert_reference}
To assess the quality of the analogies and animations in the educational context, we asked CS teachers to rate a curated set of artifacts using a set of pre-defined criteria of analogy and animation quality.

\textbf{Participants} We recruited 11 educators with experience in teaching CS, including university professors ($N=8$) and PhD candidates ($N=3$) from five institutions across three different countries. 

\textbf{Data.}
We curated a set of nine distinct topics spanning three core CS areas: \textit{Data Structures}, including Stack, Binary Search Tree (BST), and Hash Map; \textit{Algorithms}, including Recursion, Linear Search, and Merge Sort; and \textit{SE Patterns}, including Singleton, Observer, and Model--View--Controller (MVC).
Each topic was processed by \textit{ANVIL} using \texttt{gpt-4o} for the Textual and Screenplay Layer and \texttt{claude-3.7-sonnet} for the Code Layer.
Figure~\ref{fig:MVC} contains example frames extracted from a generated video.
For each topic, participants were shown (i) the target concept definition, (ii) the generated textual analogy, and (iii) the corresponding rendered animation (video).

\textbf{Evaluation Criteria.} The evaluation criteria focused on analogy and animation quality.
The \textbf{analogy quality} criteria are adapted from prior work on systematic evaluation of automatically generated educational analogies~\cite{bhavya_long-form_2024} and include: (i) \textit{Target Concept Coverage (TCC)}, assessing whether the analogy covers properties of the target concept definition; and (ii) \textit{Mapping Strength (MS)}, evaluating the logical consistency of source–target mappings.
Together, these criteria capture both the comprehensiveness of the analogy and the appropriateness of the source--target correspondences.
The \textbf{animation quality} criteria are grounded in theories of multimedia learning and motivation~\cite{mayer_multimedia_2009,clark_dual_1991} and include: (i) \textit{Alignment with Textual Analogy (ATA)}, assessing fidelity to the analogy’s mappings; and (ii) \textit{Visual Clarity (VC)}, evaluating layout and readability. Therefore, these criteria assess whether the animation is both pedagogically coherent with the textual analogy and visually effective for communication.
Each criterion was rated on a 4-point Likert scale with metric-specific anchors, omitting a neutral midpoint to encourage directional judgments~\cite{kankaras2025neutral}.

\textbf{Analysis.} We report experts' ratings using median and interquartile range (IQR) per artifact.
Inter-rater agreement was assessed using Krippendorff’s $\alpha$ on 4-point ordinal ratings. Because agreement is sensitive to score prevalence, we additionally analyse collapsed binary labels (1--2 vs.\ 3--4), and report Gwet's AC1 on these labels, which is less affected by class imbalance~\cite{ac1vsalpga}.

\textbf{Results.} \textit{Inter-rater agreement} was low on the 4-point ordinal ratings ($\alpha \le0.15$), with exact agreement between 36.8\% and 45.7\%, where most disagreements occurred between adjacent high scores (i.e., 3 vs.\ 4).
\textit{Collapsing ratings} increased exact agreement (66.9--81.0\%), but $\alpha$ remained low ($\alpha \le 0.18$) due to label prevalence.
Gwet's AC1 indicates substantial agreement for TCC (0.77), ATA (0.75), and MS (0.71), and moderate agreement for VC (0.45).
Overall, experts broadly agreed on adequacy, with disagreement concentrated in borderline cases and in VC judgments, consistent with prior work in analogy and multimedia evaluation~\cite{he_it_2022,video_agreeemtn}.
\begin{figure*}[t!]
    \centering
    \begin{subfigure}{0.49\textwidth}
        \includegraphics[width=\linewidth]{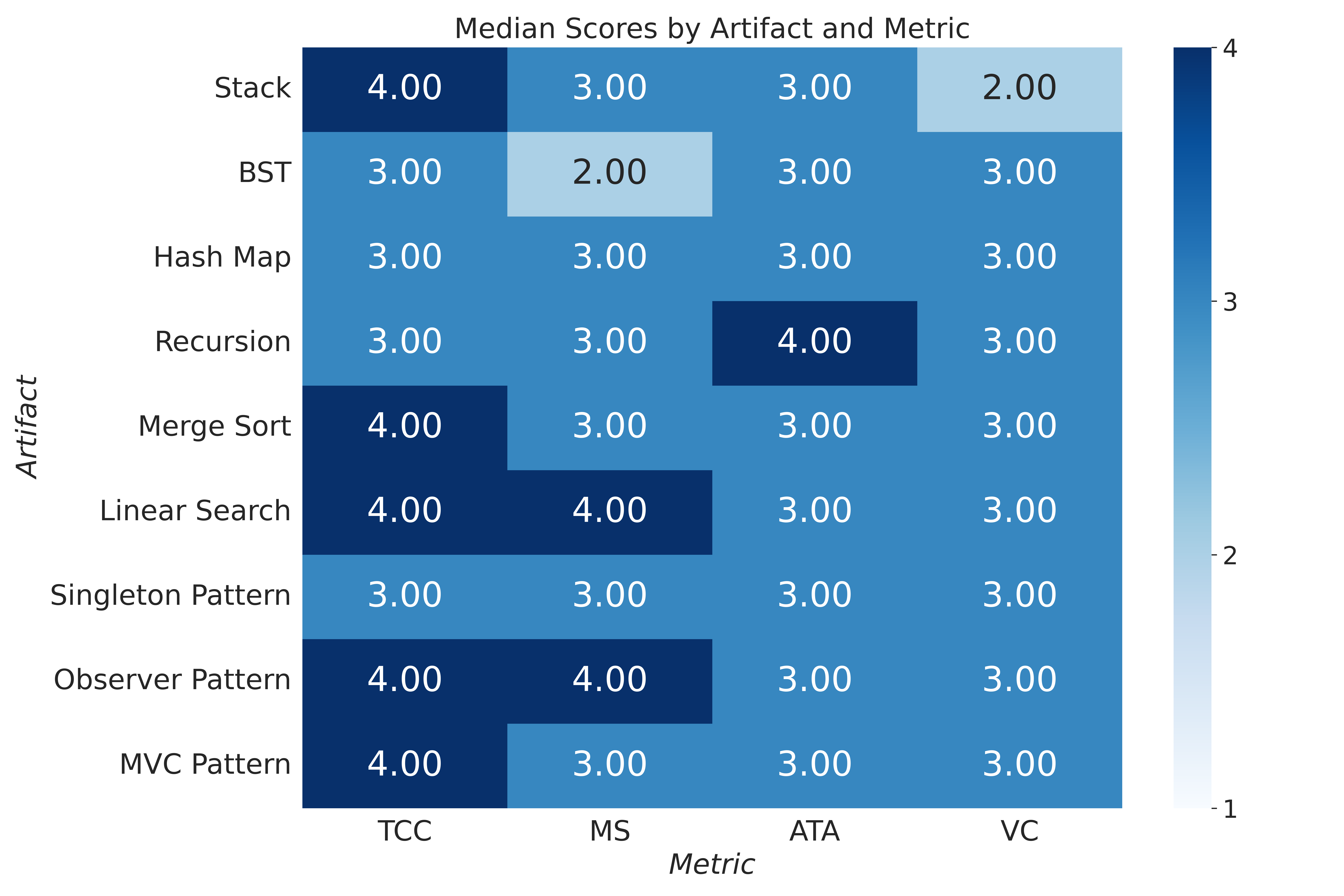}
        \caption{\small Median expert ratings. Darker colors representing more positive ratings.
        }
        \label{fig:median_heatmap}
    \end{subfigure}
    \hfill     \begin{subfigure}{0.49\textwidth}
        \includegraphics[width=\linewidth]{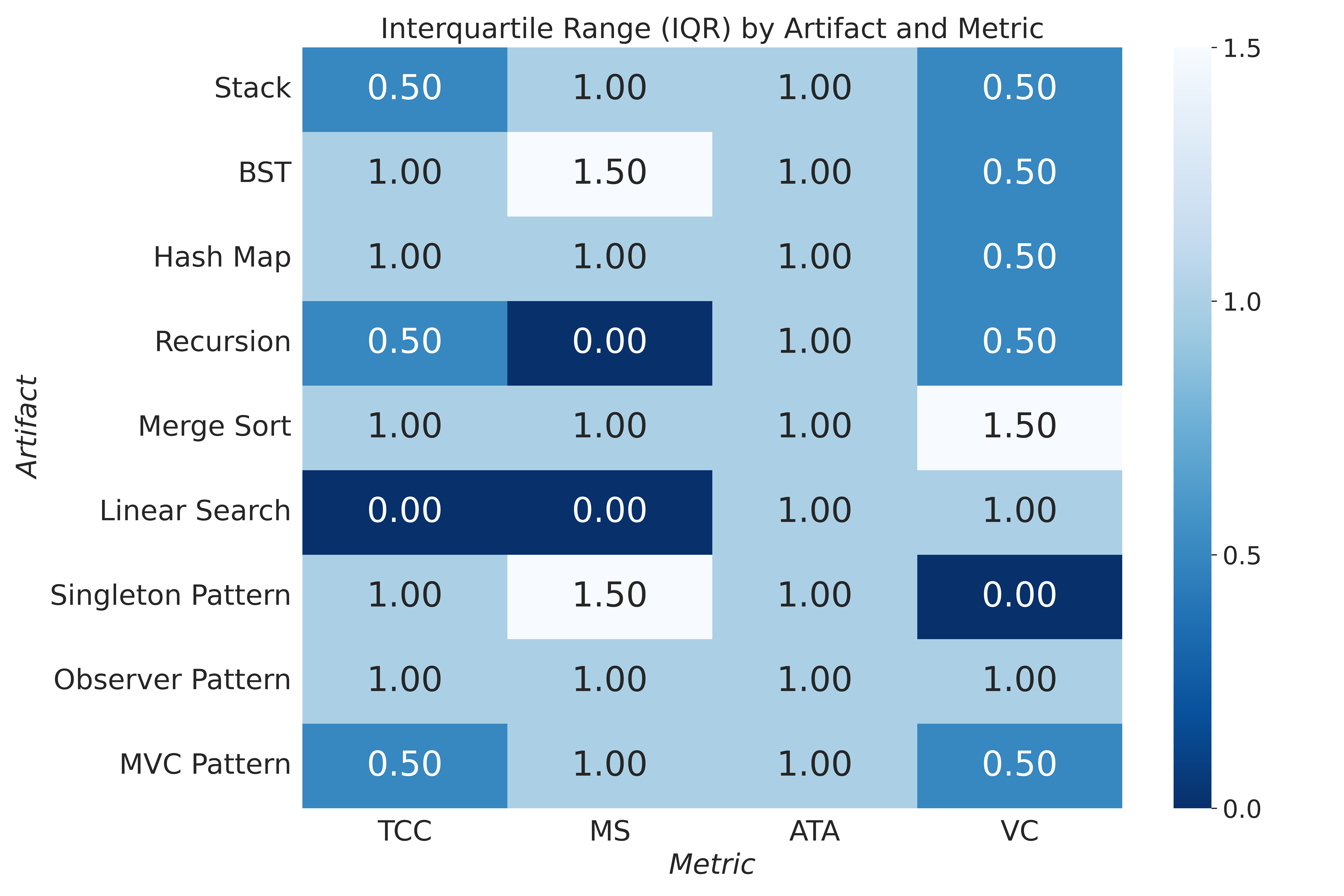}
        \caption{\small IQR of expert ratings. 
        Darker colors indicate higher expert consensus.
        }
        \label{fig:iqr_heatmap}
    \end{subfigure}
    \caption{ \small
    Heatmaps visualizing the expert evaluation results on the artifact set.}
    \label{fig:evaluation_heatmaps}
\end{figure*}
Figure~\ref{fig:evaluation_heatmaps} summarizes \textit{median} ratings and \textit{IQR} across artifacts.
Analogies were generally rated highly: TCC and MS medians were 3--4 (``Strong'' or ``Very Strong'') for most topics, with \textit{BST} as a lower-scoring case.
Animations were typically faithful to the text (high ATA medians), while VC medians were generally 3 (``Good'') with lower clarity for \textit{Stack}.
Low IQRs (usually $\le 1$) indicate moderate consensus on fundamental correctness.

\textbf{Implications for Automated Screening.}
The human evaluation results have direct implications for how automated screening should be designed.
Experts showed broad agreement on the overall adequacy of analogies and animations, with most disagreement occurring between adjacent high-scores rather than between acceptable and unacceptable outputs. Moreover, agreement was higher for analogy-related criteria than for subjective aspects of video quality. These findings motivate a modality-specific automation strategy. 

For \textbf{analogies}, experts show mostly stable agreement on adequacy, resulting in a dataset dominated by positive examples. As a consequence, an automated evaluator cannot be fully validated based on those outputs alone. 
By introducing controlled negative examples, however, we can employ a discriminative automated evaluator, whose sensitivity can be tested and validity can be demonstrated against expert judgment. 

For \textbf{animation}, experts disagree more on subjective video quality (especially VC), and it is hard to define realistic \textit{negative videos} that cover the many ways an animation can fail. Automated scoring of subjective video quality is therefore unreliable. Instead of attempting to reproduce human judgment, we can adopt a proxy-based evaluation that audits objective properties of an animation. While this proxy would not replicate the expert quality ratings, it can provide an operational signal for auditing that complements human evaluation.

\subsection{Automated Analogy Evaluation}
\label{sec:analogy_evaluation}
Following the implications from the human evaluation, we deploy an LLM-based judge~\cite{llm-judge} to automatically evaluate the quality of the analogies on TCC and MS. First, we validate the LLM-based judge against expert judgments using controlled negative examples. Second, we apply the LLM-based judge to a larger set of generated analogies to illustrate how it behaves at scale and to estimate how many system outputs would meet a baseline standard for educator use.

\textbf{Methodology.} We use \texttt{gpt-5.2} as judge of the automated evaluator. 
Given a concept definition and an analogy, the \textit{LLM-based judge} extracts the set of target properties, for which it assigns two labels on an ordinal scale: a TCC label indicating if the property is missing (1), marginally implied (2), adequately stated (3), or comprehensively covered (4), and an MS label indicating if the source-domain correspondence is misleading (1), weak (2), consistent (3), or robust (4).
The judge returns a structured \textit{JSON} record that lists the extracted properties, the assigned labels, and short textual evidence spans from the analogy.
The final aggregate TCC score is calculated as the mean property-level coverage, while the aggregate MS score is the mean mapping strength across all covered properties.
To improve \textit{scoring reliability}, we use three independent \texttt{gpt-5.2} runs and average their outputs.
Finally, we round these scores to the nearest integer on the 1–4 scale.

\textbf{Data.} The expert-rated \textit{dataset} (\Cref{sec:expert_reference}) contains predominantly positive examples. To validate whether LLM-based judge can distinguish between existing positive examples and negative examples, we generated three \textit{controlled negatives}, by following prior work that constructs plausible but incorrect analogies~\cite{parallelparc}. 
To characterize \textbf{performance at scale}, we use a \textit{dataset} consisting of $n{=}50$ concept-analogy pairs, that we already generated for the system robustness analysis (\Cref{sec:repair_stats}).

\textbf{Analysis.} We report the agreement between human raters' medians and LLM judgements. 
We analyse the TCC and MS scores of the LLM-based judge and how many generated analogies meet a baseline standard for educator use.

\textbf{Results.} On the 4-point ordinal scale, \textbf{agreement} between the LLM-based evaluator and experts was moderate ($\alpha = 0.66$ for TCC, $\alpha = 0.67$ for MS).
On the collapsed labels (1--2 vs.\ 3--4), agreement was perfect for TCC ($\alpha = 1.00$) and strong for MS ($\alpha = 0.81$).
The results show that the majority of analogies met the threshold under the collapsed two-level label (TCC: $88\%$, MS: $92\%$).
Mean scores were concentrated toward the upper end of the scale, with TCC at $3.52 \pm 0.50$ and MS at $3.55 \pm 0.47$.

\textbf{Discussion.} For analogies, the evaluator indicates that many generated candidates meet baseline adequacy criteria in terms of \textit{coverage} and \textit{mapping consistency}, which aligns with the generally positive expert judgments.
Moreover, the observed agreement between automated scores and expert labels suggests that the evaluator can help recognize analogies that meet a basic adequacy requirement for educators. This makes automated analogy evaluation suitable for surfacing likely failure cases early.

\subsection{Proxy for Video Quality}
Following the implications from the human evaluation, we propose to automate screening for video quality using a \textit{proxy} that audits whether the rendered video preserves the intended screenplay structure. 
We reconstruct a structured screenplay from the video with a Visual Language Model (VLM). We assume that if scenes, elements and actions are visible in the rendered video, they can be identified by a VLM, and can therefore be directly  compared to \textit{ANVIL}'s generated screenplay. The goal of this section is to illustrate how the audit behaves at scale.

\textbf{Rubric.}  We define the \textit{screenplay-to-video fidelity rubric} with three dimensions, each rated on a 4-point ordinal scale: (1) \textit{Scene Fidelity:} whether the video preserves the intended scene sequence and structure; (2) \textit{Element Fidelity:} whether key visual elements specified in the screenplay are present in the corresponding scenes; (3) \textit{Action Fidelity:} whether the intended actions/state changes occur and follow the screenplay ordering.

\textbf{Methodology.}
Given a rendered video, we use a \textit{VLM} (\texttt{gemini-3.0-pro}) to reconstruct an \textit{observed screenplay} from the video only.
The VLM is prompted to segment the video into scenes and, for each scene, to extract: (i) approximate timestamps, (ii) visible entities (objects/characters and key visual elements), (iii) salient actions or state changes (e.g., appear, move, fade), and (iv) any on-screen text.
The VLM returns a structured record containing a list of scenes with these fields.
To support direct comparison against \textit{ANVIL}'s \emph{target screenplay}, we prompt the VLM with the same screenplay schema used by \textit{ANVIL}.

Given a target screenplay and an observed screenplay, an \textit{LLM-based fidelity judge} aligns each target scene to the most relevant observed scene(s) and assigns rubric labels for Scene, Element, and Action Fidelity.
The judge returns a structured \textit{JSON} record that lists, for each target scene: the aligned observed segment(s), the assigned labels, and short evidence spans from the observed screenplay (objects/actions/text) supporting the decision. 
We aggregate scene-level labels by averaging across scenes, producing three video-level scores: Scene Fidelity, Element Fidelity, and Action Fidelity.
To improve scoring reliability, we use three independent \texttt{gpt-5.2} judge runs, average their outputs and round to the nearest integer on the 1--4 scale.

\textbf{Data.} We use  a set consisting of $n{=}50$ screenplay-animation pairs already generated from our system robustness analysis (\Cref{sec:repair_stats}).

\textbf{Analysis.} We report the Fidelity scores of the LLM-based judge and how many animations meet a baseline standard for educator use.

\textbf{Results.} Most videos met the threshold for \textit{Scene Fidelity} ($94\%$) and \textit{Element Fidelity} ($88\%$), whereas only $52\%$ met the threshold for \textit{Action Fidelity} under the collapsed two-level label.
Mean scores showed the same pattern: scene alignment was high ($3.44 \pm 0.67$) and element alignment was generally strong ($2.96 \pm 0.53$), but action realization was lower ($2.52 \pm 0.58$), indicating action execution and ordering as the primary failure mode.

\textbf{Discussion.} For \textit{animations}, the proxy results show that fidelity is more consistently preserved on high-level structures (scenes and elements) rather than on fine-grained actions. This difference suggests that \textit{ANVIL} reliably carries forward ``what should appear'' (objects and scene composition), while fine-grained temporal dynamics can be less clearly realized.
However, we still interpret the proxy as an audit rather than a direct measure of pedagogical quality, since low scores may also reflect limitations or errors in VLM-based reconstruction.
Nonetheless, the proxy can still serve as a useful screening signal to detect likely mismatches and guide targeted repair.

\section{User Study}
\label{sec:focus_groups}

To understand how educators would integrate \textit{ANVIL} into teaching workflows, we conducted focus groups with CS/SE educators. The goal was to gather qualitative insights about perceived value, risks, and requirements for adoption in educational settings.

\textbf{Methodology.}
\label{sec:fg_protocol}
Each of the two focus groups followed a 60-minute semi-structured protocol designed to ground discussion in concrete system outputs.
The sessions consisted of four parts: (i) an introduction of participants' prior use of analogies and visualizations in teaching; (ii) reflection on \textit{ANVIL}-generated artifacts to prompt specific critiques; (iii) a broader discussion of opportunities, challenges, and requirements for classroom integration; and (iv) a discussion on desired system features.
Sessions were audio-recorded and transcribed verbatim.

\textbf{Participants.} Nine educators from the earlier expert evaluation (\Cref{sec:expert_reference}) were divided into Focus Group A ($n=5$) and Focus Group B ($n=4$).

\textbf{Analysis.} We analysed transcripts using inductive thematic analysis~\cite{braun_using_2006}.
The {analysis was executed by (i) familiarizing with the data through repeated readings, (ii) generating initial codes directly from participant statements, and (iii) iteratively clustering codes into candidate themes.

\textbf{Results.} The analysis yielded four themes that summarize educators' perceptions of \textit{ANVIL}'s role, risks, and requirements for practical use.
~\\
\textbf{Theme 1: A creative collaborator, not a content replacement.} Across both groups, educators did not view \textit{ANVIL} as an autonomous content creator.
Its primary value was described as supporting early-stage lesson planning by proposing analogy ideas that instructors could adapt.
Several participants emphasized that the ``core value'' lies in providing an initial metaphorical framing that can serve as a lecture icebreaker and a starting point for deeper discussion.
~\\
\textbf{Theme 2: Instructor control and low-effort customization.}
Educators consistently emphasized that adoption depends on editability and instructor control.
A black-box generator that outputs only a finished video was seen as impractical: participants wanted to adjust objects, text, pacing, and terminology without needing to learn \texttt{manim}.
Multiple participants suggested interactive refinement via natural language instructions.
~\\
\textbf{Theme 3: Potential for pedagogical mismatch.}
Participants noted that, if a visualization does not preserve key constraints of the target concept, it can lead to interpretations that diverge from the intended explanation.
For example, in the \textit{Stack as a Pile of Pancakes} animation, showing the entire pile at once weakened the sense of restricted access that defines a \textit{Stack}. As a result, this property became less salient, and some operations (e.g., \textit{peek}) may appear less meaningful without additional instructor framing.
~\\
\textbf{Theme 4: Context-dependent utility and modular use.}
Educators viewed the generated videos as more suitable for supplementary or asynchronous learning than as in-lecture material, primarily due to pacing and the need to adapt narration to local terminology.
Participants expressed a strong preference for modular use of \textit{ANVIL}: using only the Textual Layer to brainstorm analogies or creating the animations starting from a traditional explanation of the concept.
\section{Limitations}
Our expert study covered a limited set of generated materials. Agreement was stronger for coarse \textit{meets baseline} judgments than for fine-grained ordinal distinctions, requiring label collapsing for threshold decisions. However, our goal was to provide an assessment of the overall quality of the generated materials, rather than a large-scale, highly granular expert evaluation.

Our automated analogy and video-fidelity measures are LLM/VLM-based methods intended for screening and larger characterization. While these measures may reflect biases of the underlying models and may not directly capture visual aesthetics or pedagogical effectiveness, they offer a scalable and efficient way to evaluate large datasets. This makes them a useful first step before presenting artifacts to teachers or students in the classroom.
\section{Conclusion and Future Work}
We presented \textit{ANVIL}, an end-to-end system that generates analogy-based instructional animations for CS topics via intermediate representations (analogy, elements, screenplay) and a repair agent that improves rendering reliability.
Our evaluation pairs educator judgments with automation designed for screening and auditing. Educators overall found the analogies and animations acceptable.
Automated analogy evaluation indicates strong performance overall and can be used for scalable candidate filtering.
The video quality proxy suggests that \textit{ANVIL} more consistently retains intended scene and element structure, while action-level dynamics are less consistently captured.
Focus groups further reveal that \textit{ANVIL}'s value lies in its role as a creative assistant, provided that instructors maintain oversight, allowing for modification to prevent misunderstandings.

From an educational perspective, \textit{ANVIL} can support the scalable creation of analogy-based instructional animations, which may help educators communicate abstract concepts through more accessible and engaging representations.

Future work will prioritize improving action correctness through stronger constraints and verification across the screenplay and code layers, and reducing instructor effort via lightweight editing and iterative refinement strategies.
Additionally, we plan to study \textit{ANVIL} in authentic teaching workflows, such as introductory programming courses.
In these settings, we will evaluate downstream impacts on student understanding of core concepts, alongside learner-experience dimensions such as engagement, clarity, and perceived helpfulness.

\section{Ethics and Consent}
Participation in this study was voluntary. All participants were informed about the study’s purpose, data collection, and data usage. Survey responses were anonymized, focus-group data were de-identified. Raw transcripts are not shared to protect confidentiality.
These study design and data processing were approved through an ethics review in the authors' institutions.

\section{Acknowledgments}
This work was conducted as part of the AI for Software Engineering (AI4SE) collaboration between JetBrains and Delft University of Technology. The authors gratefully acknowledge the financial support provided by JetBrains, which made this research possible.

\bibliographystyle{splncs04}
\bibliography{main}
\end{document}